\documentclass[epj]{svjour}
\usepackage{graphics}
\usepackage{bm}
\newcommand{\demi}{\mbox{$\frac{1}{2}$}}
\begin{document}
\title{Effective creases and contact-angles between membrane domains
with high spontaneous curvature}
\author{Jean-Baptiste Fournier\inst{1,2} \and Martine Ben Amar\inst{3}}
\institute{
  \inst{1}Mati\`ere et Syst\`emes Complexes (MSC), UMR 7057 CNRS
\& Universit\'e Paris~7, 2 place Jussieu, F-75251 Paris Cedex 05,
France\\
  \inst{2}Laboratoire de Physico-Chimie Th\'eorique, UMR 7083
  CNRS \& ESPCI, 10 rue Vauquelin, F-75231 Paris Cedex 05, France\\
  \inst{3}Laboratoire de Physique Statistique, \'Ecole Normale
Sup\'erieure,
24 rue Lhomond, F75231 Paris Cedex 05, France
}

\date{Received: date / Revised version: date}
%

\abstract{
We show that the short-scale elastic distortions that are excited  in
the vicinity of the joint between different lipidic membrane domains (at
a scale of $\simeq\!10$\,nm) may produce a ``crease" from the point of
view of the standard elastic description of membranes, i.e., an effective
discontinuity in the membrane slope at the level of Helfrich's theory.
This ``discontinuity" may be accounted for by introducing a line tension
with an effective angular dependence. We show that domains bearing
strong spontaneous curvatures, such as biological rafts, should exhibit
creases with a finite contact-angle, almost prescribed, corresponding to
a steep extremum of the line energy.  Finite contact-angles might
also occur in symmetric membranes from the recruitment of impurities at
the boundary.
} 
\titlerunning{Effective contact-angle between membrane domains}
\maketitle

\section{Introduction}

When phospholipid or like surfactants are dissolved in an aqueous
solution, they condensate into fluid bilayer membranes.
Lipid bilayers are formed by two contacting monolayers of opposite
orientations, in which the hydrophilic heads of the surfactants are
directed towards the aqueous solution while the hydrophobic tails,
confined within the sheet, are shielded from contact with
water~\cite{Israelachvili_book}. Lipid bilayer membranes are ubiquitous
in biological cells~\cite{Alberts_book}. In actual biological systems,
membranes are formed of multiple lipid or surfactant components. The
latter may laterally separate into coexisting liquid phases, or
\textit{domains}~\cite{Baumgart03,Leibler87,Seul95}, 
with distinct compositions and
distinct microscopic features, such as the membrane thickness.
Microdomains called ``rafts" are receiving increasing attention, since
they are believed to concentrate proteins that must interact with one
another to carry out important cellular
functions~\cite{Simons97,Anderson02}. 

A number of theoretical works have studied the relationship between the
curvature elasticity and line tension of domains and the resulting
formation of particular patterns or membrane
shapes~\cite{Lipowsky92,Julicher96,Allain04b,Andelman92,%
Jiang00,Kumar01,Allain04}.  In all these works, the
\textit{slope} of the membrane is assumed to be \textit{continuous} at
the domain boundary. In this paper, we argue that
although this assumption is realistic in a large number of situations, it
is not correct for domains with strongly asymmetric monolayers, as it is
often the case in the biological realm~\cite{Mouritsen_book}.

Here, we need to precise what we mean by a membrane ``slope
discontinuity" : we mean a discontinuity as far as the standard Helfrich
description of membranes is concerned. This ``discontinuity" will
obviously be resolved at a more microscopic scale, i.e., at the
molecular scale or at the scale of more involved elastic theories taking
into account short-scale structural degrees of freedom.  We do
\textit{not} refer here to apparent large-scale 
discontinuities as, e.g., the ``neck" connecting a budding domain in a
vesicle~\cite{Julicher96,Allain04b}: these are actually smooth and
continuous at a sufficiently short scale within Helfrich's theory. The
possibility of actual slope discontinuities in Helfrich's theory, or
``creases", was suggested on symmetry grounds for surfactant monolayers
in Ref.~\cite{Diamant01}, but not investigated further. Here, we
actually study the possibility of such creases in the case of
membranes.

At first sight, the condition of a continuous slope across the domain
boundary seems mathematically necessary because of the presence of a
curvature energy. Indeed, in the standard Helfrich
model~\cite{Helfrich73}, which describes the membrane as a fluid
structureless surface with a curvature elasticity, the energy density is
proportional to the square of the membrane curvature. (If the membrane is
asymmetric, there also a term linear in the curvature, with a coefficient
proportional to the asymmetry of the bilayer.)  Within this simple model,
which is essentially valid at large length-scales, if one assumes a
discontinuous membrane slope at the boundary between two domains, one
obtains a localized \textit{infinite} energy (the local energy density
diverges there as the square of a delta-function). However, this
reasoning assumes that the Helfrich model may be continuously applied
through the boundary between the domains, which is not obvious.  In
fact, as in every elastic model, there is a cutoff below which
Helfrich's model is no longer applicable. The corresponding
length-scale is comparable with a few times the membrane thickness,
i.e., $\simeq10$~nm~\cite{Goetz99}. At the \textit{joint} between
membrane domains, there are short-scale elastic degrees of freedom that
are excited: matching of the two membrane
thicknesses~\cite{Gandhavadi02,Lawrence03}, tilt of the lipidic
tails~\cite{Kuzmin05}, etc.  These degrees of freedom relax typically on
a length scale comparable with the cutoff of the Helfrich model. In this
paper, we show that these structural degrees of freedom may result in an
apparent \textit{contact-angle} in the large scale Helfrich description
(an effective jump in the membrane slope at the joint). We also
demonstrate that the energy of these extra degrees of freedom may be
``coarse-grained" into an \textit{anisotropic} line energy
$\gamma(\theta)$ for the Helfrich model, where $\theta$ is the effective
slope discontinuity at the joint.

\section{Model}
Let us consider two contacting membrane domains $\mathcal{D}_i$,
$i\in\{1,2\}$. We shall describe them more ``microscopically"
than in Helfrich's theory. Since we are interested in
the local junction between these domains, we assume
translational invariance along the $y$-direction. We take
$\mathcal{D}_1=[-L,0]$ and $\mathcal{D}_2=[0,L]$, the joint being at
$x=0$. Our model free-energy,
per unit length, is the following:
\begin{equation}
F=\sum_{i=1,2}\,\int_{\mathcal{D}_i}\!\!dx\,
\left(f_\mathrm{H}^{(i)}+f_\mathrm{str}^{(i)}+f_\mathrm{int}^{(i)}\right)
\,\,+\gamma_0\,.
\end{equation}
The first contribution $f_\mathrm{H}^{(i)}$ is similar to the bending energy
density of the Helfrich model~\cite{Helfrich73} :
\begin{equation}
f^{(i)}_\mathrm{H}=
\frac{1}{2}\kappa_i\left(\partial^2h_i\right)^2-\kappa_i\,c_i\,\partial^2h_i.
\end{equation}
Here $h_i(x)$ represents the height profile of the membrane midplane
in domain~$i$. Dealing only with small deformations, we shall neglect
everywhere terms of higher-order than $h_i^2$. With $\partial\equiv
d/dx$, the quantity $\partial^2h_i$ represents the membrane curvature,
and $\kappa_i$ and $c_i$ are the bare bending rigidities and bare
spontaneous curvatures, respectively. The latter term, which is linear
in $h_i$, is allowed by symmetry only if the membrane is dissymmetric,
i.e., if the lipid or surfactant compositions in the two monolayers are
different. For the sake of simplicity, we neglect the surface tensions
$\sigma_i$, which would give rise to a term of the form
$\frac{1}{2}\sigma_i(\partial h_i)^2$: this is reasonable as long as we
investigate length scales smaller that $(\kappa_i/\sigma_i)^{1/2}$.
Note that the Gaussian rigidities $\bar\kappa_i$~\cite{Helfrich73} can
be discarded, since in one dimension the Gaussian curvature vanishes.    

The second term, $f_\mathrm{str}^{(i)}$, arises from the
\textit{structural} energy density associated with the inner
deformations of the membranes. Although the length-scales under
consideration are comparable with a few nanometer, we keep a continuous
description. Among the many possible structural variables, two
traditional ones are the thickness of the membrane and the tilt of the lipids
relative to the membrane's midplane (see, e.g.,
\cite{Kuzmin05,Fournier99}). Note that our aim is not to discuss the
most general model, but to show, with the simplest symmetry-allowed
terms, the possible existence of the ``contact-angle" effect discussed
in the introduction. Hence, for the sake of simplicity, we consider only
the membrane thickness variable:
\begin{equation}
f_\mathrm{str}^{(i)}=
\frac{1}{2}B_i\left(u_i-a_i\right)^2+\frac{1}{2}k_i\left(\partial
u_i\right)^2.
\end{equation}
Here, $u_i(x)$ describes the thickness of the membrane in domain~$i$.
These two terms represent the free-energy cost associated with a
variation of $u_i$ with respect to its equilibrium value $a_i$, at
lowest-order in the gradient expansion~\cite{Fournier99,Huang86,Dan93}.
We expect the typical length-scale $(k_i/B_i)^{1/2}$ over which $u_i$
relaxes, i.e., the width of the joint, to be comparable with the
membrane thickness. 

There could be no possible interplay between the structural degrees of
freedom and the large-scale shape of the membrane without an energy term
coupling these quantities. We consider the following, lowest-order
interaction term:
\begin{equation}
f_\mathrm{int}^{(i)}= -\Lambda_i\left(u_i-a_i\right)\partial^2h_i.
\end{equation}
It represents the coupling between the excess thickness and the
spontaneous curvature. Like the spontaneous curvature term in
$f_\mathrm{H}^{(i)}$, this term is allowed by symmetry \textit{only} if
the two monolayers forming the membrane are dissymmetric. 

Finally, $\gamma_0$ is a line tension term that 
takes into account the microscopic interaction between the two
domains (arising, e.g., from van der Waals forces).

For the sake of simplicity, we assume $B_1=B_2\equiv B$, $k_1=k_2\equiv
k$ and $\Lambda_1=\Lambda_2\equiv\Lambda$, but we keep
$\kappa_1\ne\kappa_2$, $c_1\ne c_2$, and $a_1\ne a_2$. We shall discuss
later on the orders of magnitude of the different terms. Note that the
two-dimensional version of this model would be simply obtained by
replacing $dx$ by $d^2x$ and $\partial$ by $\bm{\nabla}$.

\subsection{Renormalized bending rigidity and spontaneous curvature}

A straightforward effect of the coupling term
$\Lambda(u_i-a_i)\partial^2h_i$ is to renormalize, away from the joint,
the bending rigidities and spontaneous curvatures~\cite{Leibler86}.
Indeed, if we neglect $\partial u_i$ (by assuming that the thickness is
uniform far from the joint), the total free-energy density reduces to
$\frac{1}{2}\kappa_i(\partial^2h_i)^2-\kappa_ic_i\partial^2h_i
+\frac{1}{2}B(u_i-a_i)^2-\Lambda(u_i-a_i)\partial^2h_i$. The latter
yields, after minimization with respect to $u_i$,
$u_i=a_i+(\Lambda/B)\partial^2h_i$, and the energy density becomes
$\frac{1}{2}\kappa'_i(\partial^2h_i)^2-\kappa'_ic'_i\partial^2h_i$,
with
\begin{eqnarray} 
\kappa'_i&=&\kappa_i-\frac{\Lambda^2}{B}\,\\
c'_i&=&\frac{\kappa_i}{\kappa'_i}\,c_i\,.
\end{eqnarray} 
These are the effective bending rigidities and spontaneous curvatures of
the membrane : the ones that would be measured in a macroscopic
experiment and the ones that should appear in the Helfrich curvature
energy.

\subsection{Equilibrium equations}

The Euler-Lagrange equilibrium equations associated with our
elastic free-energy $F$ are
\begin{eqnarray}
\label{be1}
\kappa_i\,\partial^4h_i(x)&=&\Lambda\,\partial^2u_i(x)\,,\\
\label{be2}
k\,\partial^2u_i(x)&=&B\left[u_i(x)-a_i\right]-\Lambda\,\partial^2h_i(x)\,.
\end{eqnarray}
These equations should be solved with the correct number of
boundary equations. First, there are three continuity equations at the
joint:
\begin{eqnarray}
\label{bci}
\left(h_2+\demi u_2\right)|_{x=0}&=&\left(h_1+\demi u_1\right)|_{x=0}\,\\
\left(h_2-\demi u_2\right)|_{x=0}&=&\left(h_1-\demi u_1\right)|_{x=0}\,\\
\partial h_2|_{x=0}&=&\partial h_1|_{x=0}.
\end{eqnarray}
The first two conditions represent the requirement that the polar heads
of the two membranes must match at the joint. They imply the continuity
of both $h_i$ and $u_i$. The third one, the slope continuity equation,
must be imposed in order to avoid an infinite curvature energy, as
explained in the introduction. Note that there is therefore no contact
angle at the ``microscopic" level in this model.

When calculating the variation $\delta F$ associated with the
variations $\delta u_i(x)$ and $\delta h_i(x)$, there are boundary terms
in $x=0$ in factor of $\delta u|_{x=0}$, $\delta h|_{x=0}$ and $\delta \partial
h|_{x=0}$ (remember that $u$, $h$, and $\partial h$ are continuous at
the joint). These are boundary forces and torques. The associated
equilibrium conditions are 
\begin{eqnarray}
\label{bcii}
f_1|_{x=0}+f_2|_{x=0}&=&0\,,\\
f'_1|_{x=0}+f'_2|_{x=0}&=&0\,,\\
\Omega_1|_{x=0}+\Omega_2|_{x=0}&=&0\,,
\end{eqnarray}
corresponding to the vanishing of the total force relative to $h$, the total
force relative to $u$, and the total torque relative to $h$,
respectively, with 
\begin{eqnarray}
f_i&=&-\epsilon\left(\kappa_i\,\partial^3h_i-\Lambda\,\partial u_i\right)\,,\\
f'_i&=&\epsilon\,k\,\partial u_i\,,\\
\Omega_i&=&\epsilon\left[\kappa_i\left(\partial^2h_i-c_i\right)-\Lambda\left(u_i-a_i\right)\right]\,.
\end{eqnarray}
Here, $i\in\{1,2\}$ indicates the domain where the quantity is
calculated, and $\epsilon=1$ if the quantity is calculated at the
right-hand side boundary of its domain, while $\epsilon=-1$ if it is
calculated at the left-hand side.

In addition, we assume that the domains $\mathcal{D}_i$ actually
correspond to small regions in contact with larger membrane parts,
in equilibrium under some unspecified boundary conditions and global
constraints. These outer parts transmit boundary forces
$f^\mathrm{ext}|_{x=\pm L}$, ${f'}^\mathrm{ext}|_{x=\pm L}$ and
tor\-ques $\Omega_\mathrm{ext}|_{x=\pm L}$, which yield the equilibrium
equations:
\begin{eqnarray}
f_2|_{x=L}+f^\mathrm{ext}|_{x=L}&=&0\,,\\
f_1|_{x=-L}+f^\mathrm{ext}|_{x=-L}&=&0\,,\\
f'_2|_{x=L}+{f'}^\mathrm{ext}|_{x=L}&=&0\,,\\
f'_1|_{x=-L}+{f'}^\mathrm{ext}|_{x=-L}&=&0\,,\\
\Omega_2|_{x=L}+\Omega_\mathrm{ext}|_{x=L}&=&0\,,\\
\Omega_1|_{x=-L}+\Omega^\mathrm{ext}|_{x=-L}&=&0\,.
\label{bcf}
\end{eqnarray}
These conditions can be formally derived by adding to $F$ the
contribution :
\begin{eqnarray}
F_\mathrm{ext}&=&
f^\mathrm{ext}|_{x=-L}\,h(-L)+f^\mathrm{ext}|_{x=L}\,h(L)\nonumber\\
&+&{f'}^\mathrm{ext}|_{x=-L}\,u(-L)+{f'}^\mathrm{ext}|_{x=L}\,u(L)\nonumber\\
&+&\Omega^\mathrm{ext}|_{x=-L}\,\partial h(-L)
+\Omega_\mathrm{ext}|_{x=L}\,\partial h(L),
\end{eqnarray}
and by equilibrating the boundary terms.
Notice that Eqs. (\ref{bci})--(\ref{bcf}) correctly provide 12 boundary
conditions for two integration domains subject to
differential equations globally of the sixth order.

\subsection{Equilibrium solutions}

The general solutions of the set of linear differential bulk
Eqs.~(\ref{be1})--(\ref{be2}) are: 
\begin{eqnarray}
h_i&=&A_{i1}+A_{i2}\,x+\frac{1}{2}A_{i3}\,x^2+\frac{1}{3}A_{i4}\,x^3\nonumber\\
&+&A_{i5}\,\frac{\Lambda}{\kappa_i\,q_i^2}\,e^{q_i\,x}
+A_{i6}\,\frac{\Lambda}{\kappa_i\,q_i^2}\,e^{-q_i\,x},\\
u_i&=&a_i+\frac{\Lambda}{B}\,A_{i3}
+\frac{2\Lambda}{B}\,A_{i4}\,x
+A_{i5}\,e^{q_i\,x}
+A_{i6}\,e^{-q_i\,x},\qquad
\end{eqnarray}
where $A_{i\alpha}$ are integration constants, and
\begin{equation}
q_i=\sqrt{\frac{\kappa'_i}{\kappa_i}}
\,\sqrt{\frac{B}{k}}.
\end{equation}
Since $q_i\approx\sqrt{B/k}$, we expect it to be of the order of the
inverse of the membrane thickness (unless when $\kappa'_i\to0$).

We see that the solutions are the sum of polynomials, which correspond
to the solutions one would get in a pure Helfrich model plus relaxing
exponentials, which correspond to the effect of the structural degrees
of freedom. Since these exponentials appear also in $h_i$, they may
affect the extrapolated angle at which the polynomials solutions meet
the joint. This is the source of the effect we are interested in.

\section{Detailed structure of the joint} 

To study the structure of the joint and its generic response to the
constraints transmitted by the curvature elasticity of the membrane, it
is enough to consider the situation in which two opposite far-away
torques are applied:
$\Omega^\mathrm{ext}|_{-L}=-\Omega^\mathrm{ext}|_{L}\equiv \Omega$.
Remember that those torques are acting at the
boundaries of the domains $\mathcal{D}_i$ and that they represent the
effect of the pieces of membranes that are outside the region we are
studying. We set to zero the external boundary forces:
$f^\mathrm{ext}|_{\pm L}={f'}^\mathrm{ext}|_{\pm L}=0$ (the latter would
merely transmit extra torques). We also assume that
the membrane domains
are much larger than the ``width" of the joint, i.e., $L\gg q_i^{-1}$.

\subsection{Solution of the problem}
\begin{figure}
\centerline{\resizebox{0.95\columnwidth}{!}{\includegraphics{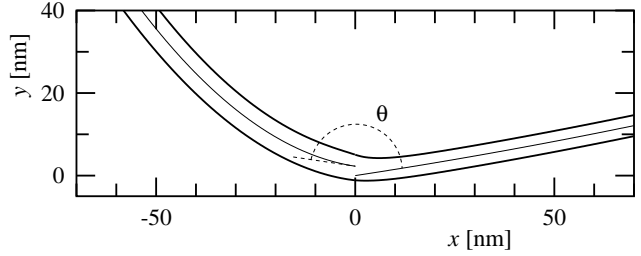}}}
\caption{Structure of the membrane in the vicinity of the joint. The
upper and lower bold lines correspond to $h_i(x)+\frac{1}{2}u_i(x)$ and
$h_i(x)-\frac{1}{2}u_i(x)$, respectively, where $i=1$ for $x<0$ and
$i=2$ for $x>0$. The parameters are
$\kappa_2=2\times10^{-19}$\,J, $\kappa_1=5\,\kappa_2$,
$c_1=(50$\,nm$)^{-1}$, $c_2=0.03\,c_1$, $B=4\times10^{15}$\,J\,m$^{-4}$,
$a_1=7$\,nm, $a_2=5$\,nm, $k=10^{-1}$\,J\,m$^{-2}$,
$\Lambda=1.2\times10^{-2}$\,J\,m$^{-2}$, and $\Omega=0$. The asymptotic
large-scale solutions $H_i(x)$ are displayed as thin lines: their
extrapolated contact-angle is $\theta=\theta_0\simeq17\,^\circ$.
For a better view, the membrane has been rotated by
$\theta/2$ counterclockwisely (and $u_i$ has been corrected in order to
yield the correct thickness when the slope is large).}
\label{joint}
\end{figure}

With the above conditions, it follows from
Eqs.~(\ref{bcii})--(\ref{bcf}) that $A_{i4}=0$,
$A_{16}=A_{25}=0$ and that
\begin{equation}
\label{curvomega}
A_{i3}=c'_i+\Omega/\kappa'_i\,.
\end{equation}
Obviously, we may choose $A_{21}=A_{22}=0$ by a simple translation and
rotation of the reference frame. Requiring that the remaining boundary
conditions be satisfied yields
\begin{eqnarray}
A_{11}&=&\frac{\Lambda\,\mu}{\kappa_1 \kappa_2}\,
\frac{\kappa_2\,q_2^3+\kappa_1\,q_1^3}{q_1^2\,q_2^2\,(q_1+q_2)}\,,\\
A_{12}&=&\frac{\Lambda\,\mu}{\kappa_1 \kappa_2}\,
\frac{B(\kappa_2-\kappa_1)}{k\,q_1\,q_2\,(q_1+q_2)}\,,
\label{A12}\\
A_{15}&=&-\mu\,\frac{q_2}{q_1+q_2}\,,\\
A_{26}&=&\mu\,\frac{q_1}{q_1+q_2}\,,
\end{eqnarray}
with
\begin{equation}
\label{mu}
\mu=a_1-a_2+\frac{\Lambda}{B}\left[
c'_1-c'_2+\left(\frac{1}{\kappa'_1}-\frac{1}{\kappa'_2}\right)\Omega
\right].
\end{equation}
We therefore find that the membrane shape $h_i$ relaxes exponentially
away from the joint with a characteristic length $q_i^{-1}$ to a
parabola of curvature $A_{i3}$, equal to the effective spontaneous
curvature $c'_i$ plus a deviation proportional to the applied external
torque (Eq.~(\ref{curvomega})). The membrane thickness, $u_i$, relaxes
exponentially to the value $a_i+(\Lambda/B)A_{i3}$, which is shifted
with respect to $a_i$ by an amount proportional to the curvature
$A_{i3}$. These features can be seen in Fig.~\ref{joint}. Away from the
joint, $h_i(x)$ tends to 
$H_i(x)=A_{i1}+A_{i2}\,x+\frac{1}{2}A_{i3}\,x^2$, as in the 
Helfrich model (for small membrane inclinations).

Although the joint essentially appears as a thickness matching, by
extrapolating the large-scale solutions $H_i(x)$ up to $x=0$, one finds
a \textit{non-zero contact-angle} (Fig.~\ref{joint}), $\theta\equiv
\partial H_1|_{x=0}-\partial H_2|_{x=0}$, given by
\begin{equation}
\theta=A_{12}\ne0\,.
\end{equation}
As expected, $\theta$ is proportional to $\Lambda$: the coupling between
the membrane shape and the structural degrees of freedom is necessary to
obtain an effective contact-angle.  Notice that there is also a mismatch
in the extrapolated membrane height, $H_1|_{x=0}-H_2|_{x=0}\ne0$, but we
may neglect it because it is smaller than the membrane thickness.

\subsection{Energy of the equilibrium configuration}

By integrating $F$ by parts and making use of the bulk equations, the total
energy at equilibrium, $F\equiv F_0$, may be expressed as boundary terms only:
\begin{eqnarray}
F_0&=&\sum_{x=0^-,-L}\frac{1}{2}\left[
f'_1\left(u_1-a_1\right)
+f_1\,h_1+\left(\Omega_1-\kappa_1\,c_1\right)\partial h_1
\right]\nonumber\\
&+&
\sum_{x=0^+,L}\frac{1}{2}\left[
f'_2\left(u_2-a_2\right)
+f_2\,h_2+\left(\Omega_2-\kappa_2\,c_2\right)\partial h_2
\right].\nonumber\\
\end{eqnarray}
Using the boundary conditions
(\ref{bci})--(\ref{bcf}), and substituting the solution
while taking the limit $Lq_i\gg1$, the total energy may be cast into the
following form:
\begin{equation}\label{formeeff}
F_0=\sum_{i=1,2}L\left[
\frac{1}{2}\kappa'_i\,A_{i3}^2-\kappa'_i\,c'_i\,A_{i3}
\right]+\gamma_0+\gamma_1\theta+\frac{1}{2}\gamma_2\,\theta^2\,,
\end{equation} 
where
\begin{eqnarray}
\label{gamma1}
\gamma_1&=&\frac{\kappa'_1\kappa'_2}{\kappa'_2-\kappa'_1}\left[
\frac{B}{\Lambda}(a_1-a_2)+c'_1-c'_2 \right]\,,\\
\gamma_2&=&-\frac{B^\frac{3}{2}}{\sqrt{k}\,\Lambda^2}
\left(\frac{\kappa'_1\kappa'_2}{\kappa'_2-\kappa'_1}\right)^2
\left(\sqrt{\frac{\kappa_1}{\kappa'_1}}
+\sqrt{\frac{\kappa_2}{\kappa'_2}}\,\right)\,,
\label{gamma2}
\end{eqnarray}
and $\gamma_0$ is the constant initially present in $F$.
One recognizes in Eq.~(\ref{formeeff}) the effective (large-scale) Helfrich
energy calculated from the asymptotic curvatures $A_{3i}$ (which are
constant here) and the renormalized bending rigidities and spontaneous
curvatures $\kappa'_i$ and $c'_i$, plus an effective line energy
$\gamma(\theta)=\gamma_1\theta+\frac{1}{2}\gamma_2\theta^2$.

Note that the large-scale curvatures $A_{3i}$'s and the effective
contact-angle $\theta$ are controlled
by the applied torque $\Omega$ (see Eqs.~(\ref{curvomega})-(\ref{mu})).
In fact, from Eqs.~(\ref{A12}) and~(\ref{mu}), one can show
with a little algebra that 
\begin{equation}
\label{eqcouples}
\Omega=-\gamma_2(\theta-\theta_0)\,,  
\end{equation} 
where $\theta_0$ is the extremum angle of $\gamma(\theta)$, given by
\begin{eqnarray}
\label{theta0}
\theta_0&=&
\Lambda\sqrt{\frac{k}{B}}\,
\frac{\kappa'_2-\kappa'_1}{\kappa'_1\kappa'_2}
\left( \sqrt{\frac{\kappa_1}{\kappa'_1}}
+\sqrt{\frac{\kappa_2}{\kappa'_2}}\,
\right)^{-1}\nonumber\\&&\times
\left[a_1-a_2+\frac{\Lambda}{B}\left(c'_1-c'_2\right)\right]
=-\frac{\gamma_1}{\gamma_2}\,.
\end{eqnarray}
Eq.~(\ref{eqcouples})
simply shows that the external torque $\Omega$ (which is transmitted up
to the joint by the bending rigidity of the membrane) is equilibrated by
the effective torque generated by the boundary line.

Note that we obtain $\gamma_2<0$. Because of this sign, the function
$\gamma(\theta)$ possesses a \textit{maximum} in $\theta=\theta_0$ and
not a minimum. This is not a problem, actually, because $\gamma(\theta)$
is only an excess energy, which does not exist by itself in the absence
of a surrounding membrane.  Hence, there is no stability criterion for
$\gamma(\theta)$ alone~\cite{stability}. The only condition physically
required is that the total energy be \textit{minimum} for $\Omega=0$, or,
equivalently (since $\theta$ is an affine function of $\Omega$) for
$\theta=\theta_0$. This is indeed the case for $L>q_i^{-1}$, since the
first term of Eq.~(\ref{formeeff}), which is obviously minimum
for $A_{3i}=c_i$, i.e., for $\Omega=0$, i.e., for $\theta=\theta_0$, is
proportional to $L$, while $\gamma(\theta)$ is not.

\section{Effective large-scale Helfrich model with a boundary crease}

We have therefore shown that our ``microscopic" model taking into
account the membrane thickness structural degree of freedom is
equivalent, \textit{at the  coarse-grained level}, to the following
(pure) Helfrich model supplemented by an anisotropic line energy for the
effective joint contact-angle $\theta\equiv\partial h_1|_{x=0}-\partial
h_2|_{x=0}$ (that must be considered as free to adjust to equilibrium):
\begin{eqnarray}
F^\mathrm{eff}&=&
\sum_{i=1,2}\,\int_{\mathcal{D}_i}\!\!dx\left[
\frac{\kappa'_i}{2}\left(\partial^2h_i\right)^2
-\kappa'_i\,c'_i\,\partial^2h_i
\right]\nonumber\\
&+&\gamma'_0+\frac{1}{2}\gamma_2\left(\theta-\theta_0\right)^2\,,
\end{eqnarray}
where $\gamma'_0=\gamma_0-\frac{1}{2}\gamma_1^2/\gamma_2>0$. 
Here, $h_i(x)$ does not describe any longer the height profile of the
membrane \textit{midplane}, it simply describes the membrane as
a whole, since at the level of Helfrich's description the whole bilayer
is treated as a pure mathematical surface. 

Let us detail the equivalence between the two models at the
coarse-grained level. The equilibrium equations and boundary conditions
at the joint, for the effective Helfrich model with energy
$F^\mathrm{eff}$, are
\begin{eqnarray}
\partial^4h_i(x)&=&0\,,\label{bulkH}\\
h_2|_{x=0}&=&h_1|_{x=0}\,,\label{bof1}\\
\partial h_2|_{x=0}&=&\partial h_1|_{x=0}+\theta\,,\label{cont1}\\
\bar f_1|_{x=0}+\bar f_2|_{x=0}&=&0\,,\label{bof2}\\
\bar\Omega_1|_{x=0}&=&-\gamma'(\theta)=-\bar\Omega_2|_{x=0}\,,
\label{torquesH}
\end{eqnarray}
with, now, $\bar f_i=-\epsilon\,\kappa_i\partial^3h_i$ and
$\bar\Omega_i=\epsilon\,\kappa_i(\partial^2h_i-c_i)$, $\epsilon$~being
defined as previously. We see indeed that since $\partial u_i$
tends to zero far from the joint, Eq.~(\ref{be1}) is equivalent at the
coarse-grained level to Eq.~(\ref{bulkH}), and that
the effective torque equilibrium equation (\ref{eqcouples})
is equivalent to Eq.~(\ref{torquesH}). The continuity Eq.~(\ref{cont1})
is postulated by the very definition of $\theta$ (which becomes a free
parameter). Finally, Eqs.~(\ref{bof1}) and~(\ref{bof2}) are only approximately
verified, but with a very good precision since
we have seen that the height mismatch in the extrapolated
large-scale solutions is smaller than the membrane thickness.

Note that Eq.~(\ref{torquesH}) shows that $\theta_0$ is the equilibrium
(effective) contact-angle of the joint, attained when no torque is
transmitted by the membrane, i.e., when the membrane is curved on both
sides of the joints at its spontaneous curvature. This corresponds to
the situation shown in Fig.~\ref{joint}.

In the limit $\Lambda\to0$ (in which $u_i$ becomes decoupled from the
membrane shape~$h_i$), we should recover the standard slope continuity
condition.  This is indeed the case, since Eqs.~(\ref{gamma2})
and~(\ref{theta0}) yield then $\theta_0\to0$ and $\gamma_2\to\infty$. We
also obtain
$\gamma'_0\to\gamma_0+\frac{1}{4}\sqrt{Bk}\left(a_1-a_2\right)^2$, which
is simply the sum of the bare line tension and the elastic energy
associated with the thickness mismatch in the absence of any coupling.

Finally, note that we have verified the equivalence of the large-scale
behavior of models $F$ and $F^\mathrm{eff}$ for arbitrary
external forcings (including external forces in addition to external
torques) by numerically solving the two boundary-value problem and
comparing the solutions far from the joint.

\section{Orders of magnitude}

Let us discuss the orders of magnitude of the various parameters for
a joint between two domains with a large thickness mismatch,
e.g., as it is the case in the junction between biological membranes and
rafts~\cite{Gandhavadi02,Lawrence03}.  For biological membranes
$\kappa_2=2\times10^{-19}$\,J and $a_2=5$\,nm are standard
values~\cite{Mouritsen_book}. Since rafts are known to be thicker and
more ordered than normal membranes, we take $a_1=6$\,nm and
$\kappa_1=5\,\kappa_2$.  From the area-stretching coefficient
$k_s\simeq0.1$\,J\,m$^{-2}$~\cite{Israelachvili_book} and the volume
conservation relation $B\simeq k_s/a_2^2$, we obtain $B\simeq
4\times10^{15}$\,J\,m$^{-4}$. To estimate $k$, we assume that the
relaxation length of $u_i$ is comparable with the membrane thickness,
which yields $k\simeq k_s$. These values are used in Fig.~\ref{joint}.

If the two monolayers are identical, we must set $c_i=0$ and $\Lambda=0$
by virtue of the resulting up-down symmetry, and hence there is no
contact-angle effect within the present model. For asymmetric membranes,
the values of the spontaneous curvature radii $c_i^{-1}$ may range
between several hundred of micrometers to a few tenth of nanometers,
depending on the difference in lipid composition or inclusion
concentrations between the two monolayers.

We must now estimate the parameter $\Lambda$. By comparing the terms
$\kappa_i\,c_i\partial^2h_i$ of $f_\mathrm{H}$ and $\Lambda
(u_i-a_i)\partial^2h_i$ of $f_\mathrm{int}$, we see that $\Lambda$
measures the dependence of the spontaneous curvature in the membrane
thickness. We expect this dependence to be significant, since it is
well-known that the spontaneous curvature of monolayers is
(qualitatively)
related with the effective conical shape of its
constituents~\cite{Israelachvili_book}. In the absence of any other
information, let us assume that a $10\,\%$ change in the membrane
thickness $a$ (which is a large variation) may change the spontaneous
curvature $c$ by an amount comparable with the spontaneous curvature
itself. This yields the estimate
$\Lambda\times0.1\,a\simeq\kappa\, c$, which, with
$\kappa\simeq5\times10^{-19}$\,J and $a\simeq5$\,nm, gives 
$\Lambda\simeq10^{-9}$\,J\,m$^{-1}\times c$.  As shown in
Table~\ref{tableau}, it follows that significant contact-angles should
only be expected for strong spontaneous curvatures. We see also that the
contribution to the effective line tension depends very weakly on
$\Lambda$. 

\subsection{Rigidity of the crease}

To determine the angular rigidity of the joint, we can estimate the
curvature radius $R^\star$ which must be applied to the joint in order
to obtain an angular variation of $0.1$\,rad ($\simeq5.7$\,deg). From
Eq.~(\ref{eqcouples}), we obtain
$R^\star\simeq\kappa/(\gamma_2\times.1)$ (see Table~\ref{tableau}). We
see that even for large values of $\Lambda$, $R^\star$ is
quite microscopic; hence, the joint appears as
extremely rigid. In consequence, we may almost view $\theta_0$ as a
\textit{prescribed} contact-angle.

\begin{table}
\caption{Values of the asymmetry-dependent constant $\Lambda$
and the corresponding joint parameters as a function of the
membrane spontaneous curvature radius $c^{-1}$.}
\label{tableau}
\begin{center}
\begin{tabular}{c|cccc}
$c^{-1}$~[nm] & $\Lambda$~[pN/nm] & $|\theta_0|$~[deg]  & 
$\gamma'_0$~[pN] & $R^\star$~[nm] \\
\hline\hline
50&20&23&$\gamma_0$+5.002&96\\
300&3.3&2&$\gamma_0$+5.001&0.7\\
$10^4$&$0.01$&$6\times10^{-3}$&$\gamma_0$+5&$6\times10^{-6}$\\
0&0&0&$\gamma_0$+5&0
\end{tabular}
\end{center}
\end{table}

\section{Discussion and conclusions}

The present study shows that if the structural distortions
arising within the joint between membrane domains are \textit{coupled}
with the membrane shape, then one should generically expect, in the
large-scale elastic description, an effective slope
discontinuity~$\theta$.  The boundary line energy
should then be considered, in general, a function
$\gamma(\theta)$. 

It should be noted, however, that if an extremely sharp extremum occurs
in $\gamma(\theta)$ for $\theta_0=0$, then one effectively recovers the
usual slope continuity condition and one should not bother about the
angular dependence of $\gamma(\theta)$.  The situation is therefore
subtle and it is essential to consider the conditions for which either
$\theta_0\ne0$ or the extremum of $\gamma(\theta)$ is shallow (compared
to the torques produced by macroscopic elastic distortions).

It is not likely that the extremum of $\gamma(\theta)$ be effectively
shallow. Indeed, finite apparent angular discontinuities must be
resolved within the joint by strong continuous distortions, and
the latter will always cost energies very large compared to those needed to
curve the membrane on a macroscopic scale. It is thus more promising to
look for $\theta_0\ne0$, but this is possible only for membranes with
broken up-down symmetry, hence membranes bearing a spontaneous
curvature. This is one of the reasons for the choice of the specific
structural model studied here.

In fact, in this paper, we have examined the consequences of the most
elementary boundary distortion: the membrane thickness matching
(occurring, e.g., at a raft boundary). For the sake of simplicity, we
have considered homogeneous membranes and we have assumed a
\textit{linear} coupling between the thickness and the shape of the
membrane (the latter requiring the broken up-down symmetry).  We have
demonstrated, within this model, the existence of an effective
anisotropic line tension $\gamma(\theta)$ and we have shown that it has
such a sharp extremum $\theta_0$ that the latter can be considered
almost as a prescribed \textit{contact-angle}.  We have obtained
$\theta_0\simeq0$ for membranes with macroscopic curvature radii, but
$\theta_0$ finite, for membranes with microscopic curvature radii.  This
is indeed interesting, because biological membranes, and especially
rafts are known to be very asymmetric, with sphingomyelin and
glycosphingolipids enriched in the exoplasmic leaflet and glycerolipids
in the cytoplasmic leaflet~\cite{Simons97}.  Curvature radii of a few
tenth of nanometers, as in Table~\ref{tableau}, are thus plausible, and
biological rafts could be expected to have contact-angles as large as 
$10$---$20^\circ$. 

One should bear in mind that we have used a continuous model to
investigate the effect of nanometric scale distortions. This is actually
the limit of validity of continuous membrane models, and one should not
expect better than semi-quantitative results. Besides, there are many
other non-linear terms that might be taken into account, and also many
other structural degrees of freedom. Real contact-angle might therefore
be significantly stronger, or smaller, than what we have estimated here;
experiments, or numerical simulations (e.g., with coarse-grained
molecules) should be the best way to give a quantitative answer. Note
also that we have neglected, for the sake of simplicity, the 
higher-order terms proportional to $\partial^2u$ and $(\partial^2u)^2$
which have been investigated in~\cite{Kuzmin05}: they certainly would
renormalize our results but cannot, for symmetry reasons, be the direct
source of an effective contact-angle.

An interesting effect that we have neglected here is the possibility
that the boundary will recruit dilute impurities and accumulate them in
the joint, in order to relax some of the microscopic distortions. Since
such impurities will in general have different concentrations in the two
monolayers, they will locally enhance, within the joint, the local
spontaneous curvatures and thus the local coupling constant $\Lambda$,
yielding again finite contact-angles (even for up-down symmetric host
membranes).  These impurities could also soften the elasticity of the
boundary distortions.

\end{document}